# Toggle Spin-Orbit Torque MRAM with Perpendicular Magnetic Anisotropy

Naimul Hassan[1], Susana P. Lainez-Garcia[1], Felipe Garcia-Sanchez[2,3], Joseph S. Friedman[1]

[1]Department of Electrical & Computer Engineering
The University of Texas at Dallas
800 W. Campbell Rd.
Richardson, TX, 75080
U.S.A.

[2]Istituto Nazionale di Ricerca Metrologica
Strada delle Cacce, 91
10135 Torino
Italy

[3]Departamento de Física Aplicada
Universidad de Salamanca
Plaza de la Merced s/n.
37008 Salamanca
Spain

Correspondence should be addressed to joseph.friedman@utdallas.edu.

**Spin-orbit torque (SOT) is a promising switching mechanism for magnetic random-access memory (MRAM) as a result of the potential for improved switching speed and energy-efficiency. It is of particular interest to develop an SOT-MRAM device with perpendicular magnetic anisotropy (PMA) in order to leverage the greater density and thermal stability achievable with PMA as opposed to in-plane magnetic anisotropy. However, the orthogonality between SOT and PMA prevents deterministic directional switching without an additional device component that breaks the symmetry, such as an external magnetic field or complex physical structure; not only do these components complicate fabrication, they also are not robust to variations in fabrication and applied switching current. This letter therefore proposes a simple SOT-MRAM structure with PMA in which deterministic toggle switching is achieved without requiring additional device components. Furthermore, this toggle PMA SOT-MRAM is shown to be far more robust than previous approaches for directional PMA SOT-MRAM, with greater than 50% tolerance to applied switching current magnitude. This letter describes the physical structure and toggle switching mechanism, provides micromagnetic simulations demonstrating its feasibility, and evaluates the robustness and tolerance to material parameters to guide the fabrication of optimized devices that will jumpstart the third generation of MRAM.**

Magnetic random-access memory (MRAM) is a promising candidate for next-generation data storage due to its non-volatility[1], high speed, and energy efficiency[1–3]. The core of each MRAM bit cell is composed of a magnetic tunnel junction (MTJ) that can be switched between two resistance states, and this MTJ is accompanied by complementary circuitry to read and write the magnetic state. Following the development of MRAM switching driven by a magnetic field, spin-transfer torque (STT)-MRAM has become preeminent due to its increased density and energy efficiency. In particular, STT-MRAM with perpendicular magnetic anisotropy (PMA) is preferred over in-plane anisotropy due to its higher density[5] and increased thermal stability, which results in a longer data retention time. However, STT-MRAM has several limitations resulting from sharing the read and write path, including degradation of the tunnel barrier from repeated switching. Therefore, spin-orbit torque (SOT) switching has recently been developed in order to overcome the limitations of STT by decoupling the write current path from the MTJ tunnel barrier.

However, SOT produces a spin current polarized in the in-plane direction, which cannot switch an MTJ with PMA. Several approaches have recently been developed to break the SOT symmetry, thereby enabling SOT-MRAM with PMA: one approach is to apply an in-plane magnetic field along the direction of the writing current[4]; another approach involves the deformation of the structure[6,7]; a third requires tilting of the anisotropy by wedge-shaped ferromagnets[5]; another uses an antiferromagnet–ferromagnet bilayer system[9], a fifth uses competing spin currents[10]. Unfortunately, all of these approaches increase the fabrication complexity, are highly sensitive to the switching current duration and magnitude, or increase the switching energy. It is therefore critical to develop an energy-efficient PMA SOT-MRAM that is simple to fabricate, robust to switching current parameters, and does not require an external magnetic field.

We therefore propose toggle PMA SOT-MRAM that exploits the precessional nature of field-like SOT to achieve field-free and energy-efficient switching with a simple structure that is robust to the switching current magnitude and duration. Leveraging the SOT toggle switching suggested by Legrand *et al.*[11], we apply unidirectional SOT current pulses that toggle the PMA

MRAM between the parallel and anti-parallel states. With this toggle switching, each SOT pulse flips the stored magnetization irrespective of its initial direction; the write circuit can use this toggle switching mechanism for selective directional switching[13,14]. This toggle switching is in contrast to the bidirectional currents required for conventional SOT-MRAM devices with directional switching, and is analogous to Savtchenko toggle switching of commercially-available field-switched MRAM[12]. This toggle SOT-MRAM device is highly robust to the switching current magnitude and duration, thus simplifying the write circuit and improving system efficiency. In particular, this switching phenomenon is shown here to tolerate write current imprecision greater than 50% and rise times slower than 200 ps. Furthermore, the device structure consists of a minimal number of planar layers, thereby simplifying fabrication and increasing the potential for continued MRAM scaling. The proposed memory device thus provides the first robust approach to simultaneously leverage the energy-efficiency of SOT and the thermal stability of PMA without requiring complex fabrication or an external magnetic field.

The structure of the SOT-driven toggle PMA MRAM is shown in Fig. 1 as a three-terminal magnetic tunnel junction[15] composed of a heavy metal, free ferromagnet, insulating tunnel barrier, fixed ferromagnet, and compensating ferromagnet. Current through the heavy metal induces SOT on the adjacent free ferromagnet, while the compensating ferromagnet cancels the stray field. Both the free and fixed ferromagnets have PMA, with a $\hat{z}$-directed easy axis and hard x-y plane. The magnetization of the fixed ferromagnet is in the $-\hat{z}$ direction through antiferromagnetic coupling, while the free ferromagnet can toggle between stable relaxed states in the $+\hat{z}$ and $-\hat{z}$ directions by applying a unidirectional current of a certain range through the write path. The resistance state of the MTJ can be determined through the tunneling magnetoresistance effect with a small current passed through the read path.

This magnetization switching mechanism can be understood as follows for an initial low magnetoresistance state where both ferromagnet magnetizations are stable in the $-\hat{z}$ direction (Fig. 2a). When a write current is applied through the heavy metal in the $+\hat{y}$ direction, a $+\hat{z}$-directed SOT spin current is produced that is polarized in the $-\hat{x}$-direction. The interplay between the PMA field ($\mu_0 H_K \hat{z}$) and the field-like ($H_T$) and damping-like ($H_L$) components of the SOT causes the magnetization $\hat{m}$ of the free ferromagnet to precess around the net magnetic field of the system, $\vec{B}_{net}$ (Fig. 2b). $\hat{m}$ crosses the hard x-y plane as it precesses (Fig. 2c), and eventually reaches a stable state with positive z magnetization not aligned with the easy axis (Fig. 2d). Once the SOT current excitation is removed, $\hat{m}$ relaxes to the nearest position along the easy axis in the $+\hat{z}$ direction, switching the MTJ to a high magnetoresistance state (Fig. 2e). When the next write current is applied, the magnetization of the free ferromagnet toggles back to the $-\hat{z}$ direction; as $\hat{m}$ crosses the hard x-y plane exactly once during each application of an SOT current, the MTJ toggles between the parallel and anti-parallel states every time an SOT current is applied.

This toggle switching is demonstrated via the micromagnetic simulations[16] of Fig. 3. The circular monodomain free ferromagnet has 30 nm diameter and 1.2 nm thickness, and the material parameters are taken from Zhang *et al.*[17] As shown in Figs. 3a-b, the magnetization is initialized at $m_z = -1$, and the application of the SOT current causes $\hat{m}$ to cross the hard axis and reach a stable excited state with $m_z = +0.4$. When the SOT current is removed, $\hat{m}$ precesses around the easy axis as it relaxes to $m_z = +1$, having flipped its orientation relative to the easy axis. Repeated SOT current pulses cause this toggle MRAM to switch magnetization states with each pulse, as demonstrated in Fig. 3c with four consecutive SOT pulses of 4 ns duration with 10 ns of relaxation between each.

This toggle MRAM device is promising for the next generation of non-volatile memory due to its simplicity and robustness to input excitation. To demonstrate the exceptional robustness of this toggle MRAM switching, micromagnetic simulations were performed to determine the sensitivity of the switching process on the current amplitude and dynamics. Furthermore, our results provide material design guidelines to maximize the robustness of the switching phenomenon.

To evaluate this robustness – and therefore the precision required to design a CMOS driver circuit – we define the toggle range within which the switching mechanism behaves properly as

$$\text{Toggle Range} = \mu_0 H_{L,max} - \mu_0 H_{L,min}, \qquad (1)$$

where $\mu_0 H_{L,min}$ and $\mu_0 H_{L,max}$ denote the maximum and minimum damping-like SOT fields, respectively, for which the toggle switching proceeds properly. As can be seen in Fig. 4a, damping-like SOT fields smaller than $\mu_0 H_{L,min}$ are insufficient to cause the free ferromagnet magnetization to cross the hard axis; therefore, no switching occurs. For damping-like SOT fields greater than $\mu_0 H_{L,max}$, the excited stable state is so close to the hard axis that thermal noise can cause $\hat{m}$ to relax in an unpredictable manner, which is highly problematic for memory applications. In particular, we use a threshold of $m_z = |0.2|$ which is stricter than the 0.15 value used by Torrejon *et al.*[18] and ensures relaxation in less than 10 ns. For a robustness metric more relevant to device and circuit fabrication, we further define the toggle range ratio as

$$\text{Toggle Range Ratio} = \frac{\mu_0 H_{L,max} - \mu_0 H_{L,min}}{\mu_0 H_{L,min}} = \frac{J_{SOT,max} - J_{SOT,min}}{J_{SOT,min}}, \qquad (2)$$

where $J_{SOT,min}$ and $J_{SOT,max}$ denote the actual maximum and minimum in-plane current values through the heavy metal layer corresponding to proper toggle switching.

Based on these metrics, the robustness to the magnitude and rise time of the SOT current pulse is analyzed for various ratios between the transverse magnetic field $\mu_0 H_T \hat{\sigma}$ and longitudinal magnetic field $\mu_0 H_L (\hat{m} \times \hat{\sigma})$, expressed as the field-to-damping component ratio $\beta$ as a function of the spin polarization direction of SOT current, $\hat{\sigma}$. $\beta$ values ranging from 2 to 8 have been reported[19], with Legrand[11] having demonstrated toggle switching for $\beta$ values between 1.82 and 4.35. As shown in Fig. 4b, the toggle range is maximized in the $\beta$ range explored by Legrand[11], with the toggle range ratio above 50% for $\beta$ between 4 and 5 for step current pulses with zero rise time. The toggle range and ratio decay with increased rise time as shown in Figs. 4c-d, though proper toggle switching persists for rise times greater than 200 ps. Importantly, it is observed that the toggle range ratio for $\beta = 4$ remains greater than 50% for rise time less than 50 ps; this ratio is significantly larger than has previously been demonstrated with directional SOT switching in a deformation free antiferromagnet–ferromagnet bilayer system by an energy efficient non-competing current[9]. It should also be noted that while large $\beta$ values do not provide a particularly large toggle range and ratio in response to step inputs, large $\beta$ values provide the greatest toggle range and ratio for SOT current pulses with a large rise time.

In conclusion, toggle switching is a simple and effective approach for SOT-MRAM with PMA, and provides increased robustness than directional switching. The use of toggle switching and the increased robustness both reduce the hardware overhead of the write circuits, and the simplified device structure further reduces the area and improves the energy efficiency of MRAM caches. This proposed toggle MRAM device therefore provides a promising pathway for the incorporation of PMA devices in a new generation of compact, highly-efficient, and robust MRAM with SOT switching.

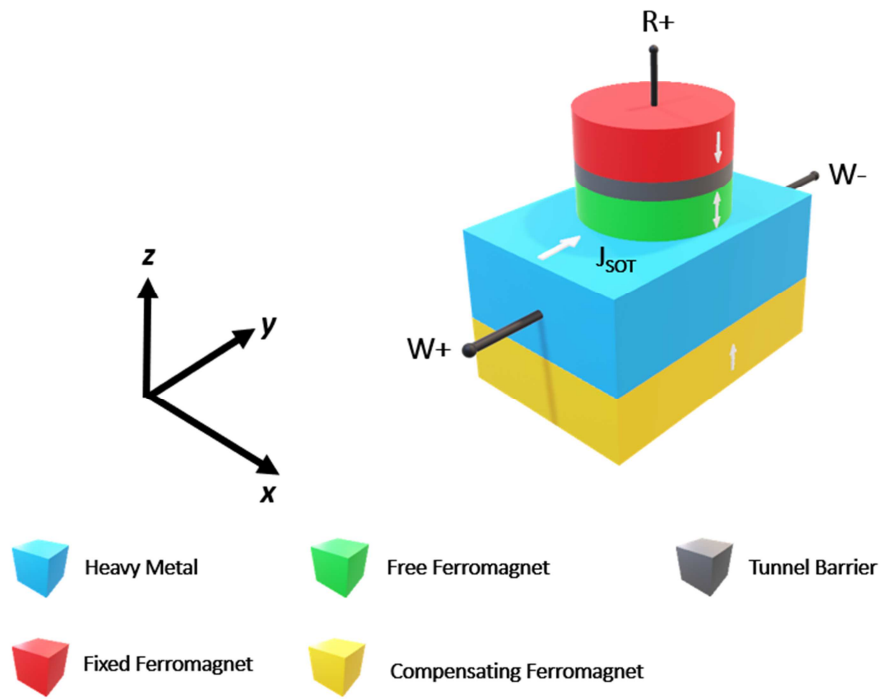

Fig. 1. Schematic of the three-terminal MTJ for toggle MRAM, with state recorded in the free layer and a compensating ferromagnet for stray field cancellation. Write current from W+ to W- creates SOT that switches the magnetization of the free ferromagnet; the state of the MTJ can be read by applying a voltage between R+ and W-.

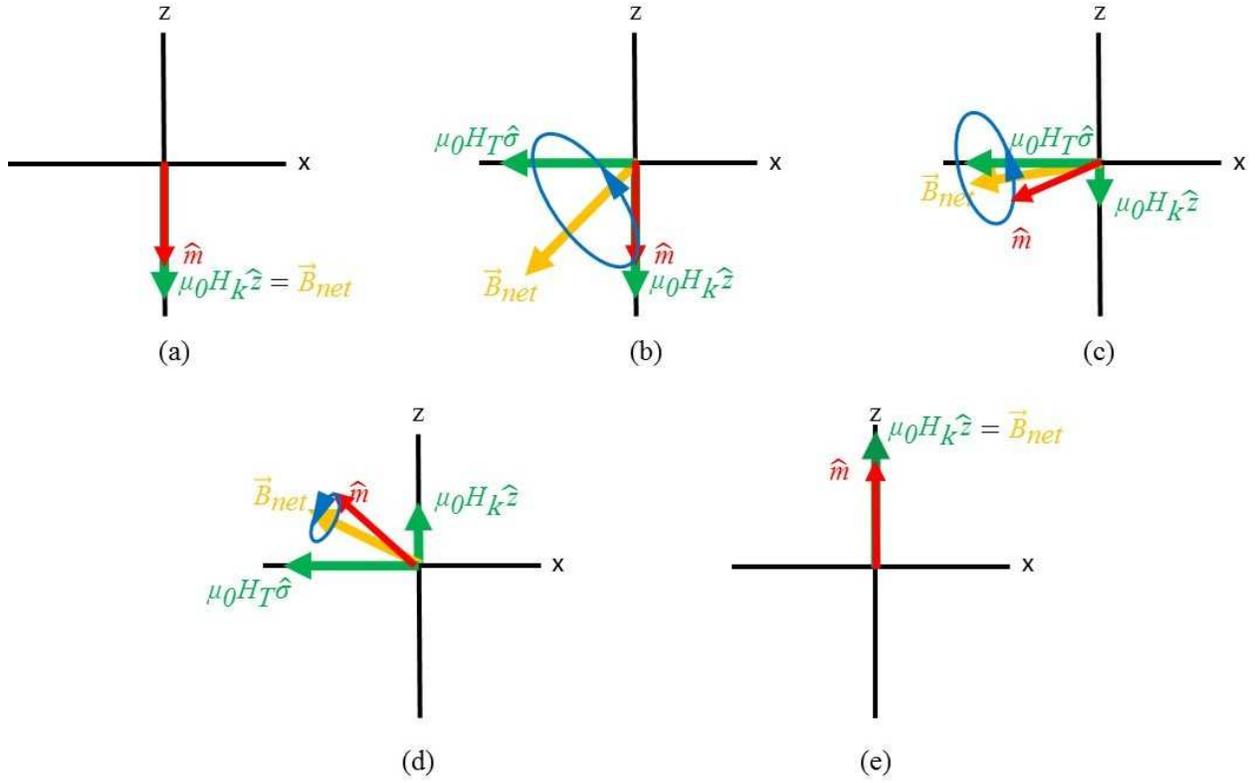

Fig. 2. Toggle switching process. (a) Free layer magnetization $\hat{m}$ is initially relaxed at $-\hat{z}$ direction before the application of SOT excitation. (b) Application of field-like SOT, $\mu_0 H_T \hat{\sigma}$, causes $\hat{m}$ to precess around $\vec{B}_{net}$. (c) Gilbert damping decreases the radius of the circular trajectory and applies a torque on $\hat{m}$ to align it along $\vec{B}_{net}$. Meanwhile, the magnitude and direction of $\vec{B}_{net}$ also changes due to the varying perpendicular magnetic anisotropy field $\mu_0 H_{K,eff} m_z \hat{z} = \mu_0 H_K \hat{z}$. (d) The precessional and damping forces cause both $\hat{m}$ and $\vec{B}_{net}$ to cross the hard x-y plane. The radius of the circular trajectory approaches zero, causing $\hat{m}$ to reach a stable excited state. (e) Switching off the SOT excitation causes $\hat{m}$ to relax in the $+\hat{z}$ direction (nearest easy axis direction), opposite the initial state.

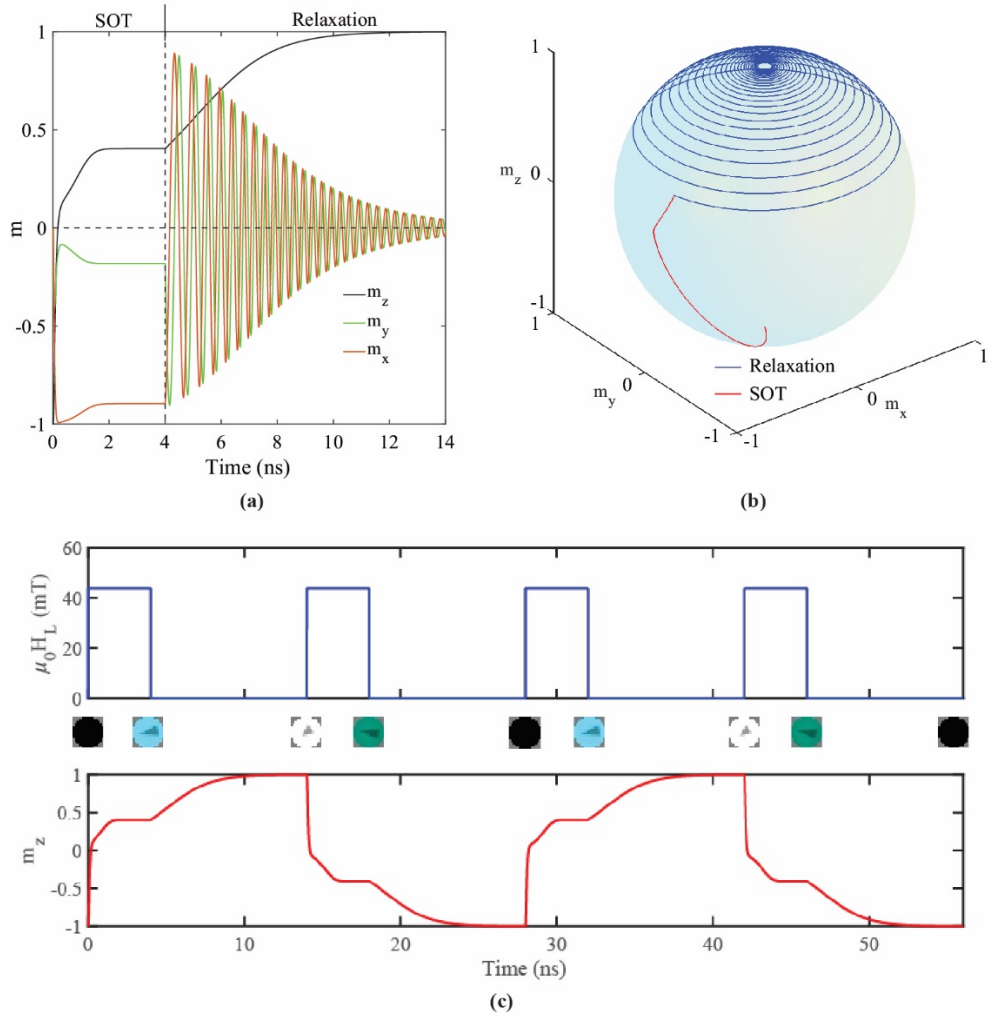

Fig. 3. Micromagnetic simulation of toggle switching. (a) Time vs. normalized magnetization. We inject an SOT current for 4 ns, and then let the system relax in the absence of current. $\hat{m}$ is initially in the -$\hat{z}$ direction, crosses the hard axis at $t = $ ~180 ps, and reaches a stable excited state of $m_z = 0.4$ at $t = $ ~2 ns. After the current is switched off at $t = 4$ ns, $\hat{m}$ relaxes along the +$\hat{z}$ direction. (b) Trajectory of $\hat{m}$ in a unit sphere during SOT excitation (red) and relaxation (blue). (c) Multiple 4 ns SOT currents are applied with magnetic field $\mu_0 H_L$, causing $\hat{m}_z$ to toggle with each applied SOT current pulse, as depicted via micromagnetic simulation screenshots that demonstrate switching between the +$\hat{z}$ (white) and −$\hat{z}$ (black) directions.

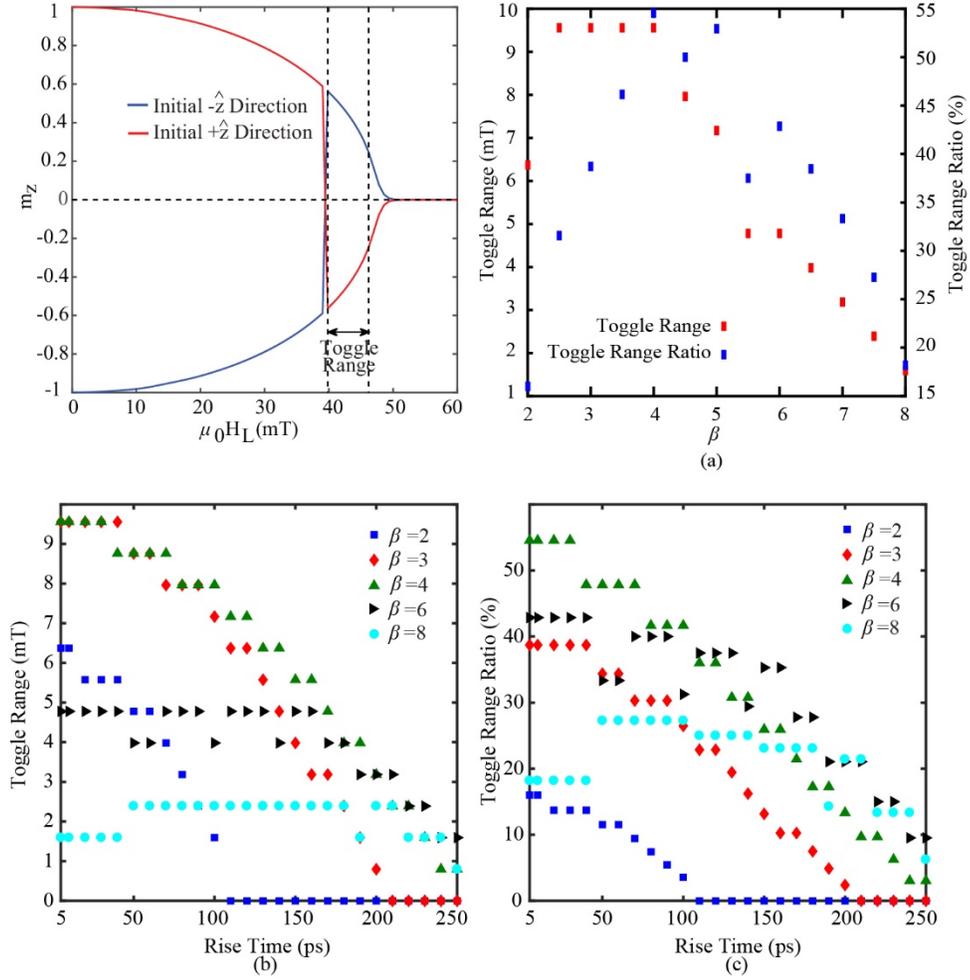

Fig. 4. Sensitivity of the toggle switching to variations in damping-like SOT field strength, rise time, and $\beta$. (a) Stable excited state magnetization as a function of damping-like SOT field $\mu_0 H_L$. The dotted vertical lines denote the boundary of the toggle range for magnetization threshold $|m_{z,th}| = 0.2$. (b) $\Delta\mu_0 H_L$ and toggle range ratio as a function of $\beta$ for a step damping-like SOT excitation (zero rise time). (c) Toggle range as a function of rise time for various $\beta$ values, illustrating that the damping-like SOT field strength becomes less sensitive to rise time for higher $\beta$ values. (d) Rise time vs. toggle range ratio for various $\beta$ values.